\documentclass[pdflatex,referee,sn-nature]{sn-jnl}% Style for submissions to Nature Portfolio journals
\usepackage{graphicx}%
\usepackage{multirow}%
\usepackage{amsmath,amssymb,amsfonts}%
\usepackage{amsthm}%
\usepackage[title]{appendix}%
\usepackage{xcolor}%
\usepackage{textcomp}%
\usepackage{manyfoot}%
\usepackage{booktabs}%

\theoremstyle{thmstyleone}%
\theoremstyle{thmstyletwo}%

\theoremstyle{thmstylethree}%

\AtBeginDocument{\raggedbottom}%%\raggedbottom
\begin{document}

\title[Phonon-driven Floquet-Bloch states probed by quantum beat spectroscopy]{%
    Phonon-driven Floquet-Bloch states probed by quantum beat spectroscopy
}


\author[1,2]{\fnm{Yu-Chan} \sur{Tai}}
\author[2,3,4]{\fnm{Chih-Wei} \sur{Luo}}
\author[5]{\fnm{Noriaki} \sur{Takagi}}
\author[6]{\fnm{Hiroshi} \sur{Ishida}}
\author[2,3]{\fnm{Chun-Liang} \sur{Lin}}
\author*[1]{\fnm{Ryuichi} \sur{Arafune}}\email{ARAFUNE.Ryuichi@nims.go.jp}

\affil*[1]{%
    \orgdiv{Research Center for Materials Nanoarchitectonics}, %
    \orgname{National Institute for Materials Science}, %
    \orgaddress{\street{1-1 Namiki}, \city{Tsukuba}, \postcode{305-0044}, \state{Ibaraki}, \country{Japan}}%
}

\affil[2]{%
    \orgdiv{Department of Electrophysics}, %
    \orgname{National Yang Ming Chiao Tung University}, %
    \orgaddress{\street{No. 1001, Daxue Rd.}, \city{East Dist.}, \postcode{300093}, \state{Hsinchu City}, \country{Taiwan}}%
}

\affil[3]{%
    \orgdiv{Institute of Physics and Center for Emergent Functional Matter Science}, %
    \orgname{National Yang Ming Chiao Tung University}, %
    \orgaddress{\street{No. 1001, Daxue Rd.}, \city{East Dist.}, \postcode{300093}, \state{Hsinchu City}, \country{Taiwan}}%
}

\affil[4]{%
    \orgname{National Synchrotron Radiation Research Center}, %
    \orgaddress{\street{101 Hsin-Ann Road}, \city{Hsinchu Science Park}, \postcode{300092}, \state{Hsinchu City}, \country{Taiwan}}%
}

\affil[5]{%
    \orgdiv{Graduate School of Human and Environmental Studies}, %
    \orgname{Kyoto University}, %
    \orgaddress{\street{Yoshida-nihonmatsu-cho}, \city{Sakyo-ku}, \postcode{606-8501}, \state{Kyoto}, \country{Japan}}%
}

\affil[6]{%
    \orgdiv{College of Humanities and Sciences}, %
    \orgname{Nihon University}, %
    \orgaddress{\street{3-25-40 Sakurajosui}, \city{Setagaya-Ku}, \postcode{156-8550}, \state{Tokyo}, \country{Japan}}%
}


\abstract{%
Controlling material excitations offers access to novel fundamental and technological properties.
The paradigm of Floquet engineering, the manipulation of the electronic structure using a coherent and time-periodic driving source, has attracted significant attention.
While most realizations rely on strong optical fields, coherent phonons provide an alternative route to realizing Floquet-Bloch states, and are expected to enable substantially longer-lived Floquet-Bloch states.
We show that laser-excited coherent phonons drive Floquet-Bloch states.
Using time-resolved multiphoton photoemission combined with quantum beat spectroscopy on graphene-covered Ir(111), we track the coherent electronic dynamics of the image-potential states dressed by coherent phonons.
The beat signal indicates the presence of sideband structure with the coherent-phonon frequency as its fundamental period, consistent with Floquet theory.
Furthermore, an independent oscillation in intensity at the same frequency was observed, confirming excitation of the coherent phonon mode.
Compared with conventional light-driven Floquet-Bloch states, the observed phonon-driven Floquet-Bloch states persist for one to two orders of magnitude longer.
These results establish a time-domain route to identifying phonon-driven Floquet-Bloch states and reveal their formation on ultrafast timescales.
}

\maketitle

\begin{figure}[!htbp]
	\centering
	\includegraphics[width=1\textwidth]{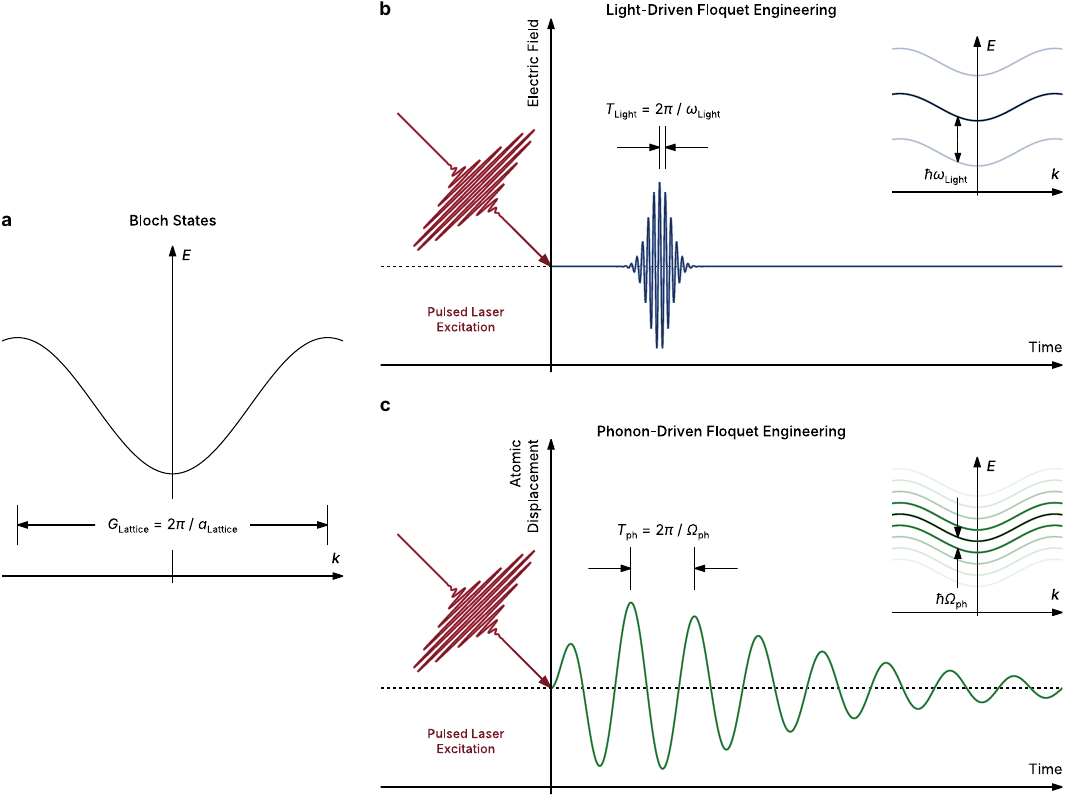}
    \begingroup
    \setstretch{1.2}
	\caption{\small\textbf{Schematics of light- and phonon-driven Floquet engineering.}
        \textbf{a,} Bloch states arise from spatial translational symmetry of the crystal lattice.
        When an additional temporal periodicity is imposed, the Bloch states evolve into Floquet-Bloch states.
        \textbf{b,} In light-driven Floquet engineering, an optical field provides a temporal periodicity $T_{\mathrm{light}}$, producing Floquet sidebands separated by $\hbar\omega_{\mathrm{light}}$.
        These states are typically confined to the laser-pulse duration ($\sim 100$~fs).
        \textbf{c,} In phonon-driven Floquet engineering, a coherent phonon provides the temporal periodicity $T_{\mathrm{ph}}$, producing sidebands separated by the phonon energy $\hbar\Omega_{\mathrm{ph}}$.
        Since the drive is sustained by coherent lattice motion, the resulting Floquet-Bloch states can persist beyond the optical-pulse duration.
    } \label{fig01}
    \endgroup
\end{figure}

External control of periodicity has become a central strategy for engineering emergent quantum phases in condensed matter systems.
Whereas conventional periodicity engineering acts in space through structures such as moir\'{e} superlattices~\cite{Nature.632.1032.2024,Dean_Nature,Cao_Nature}, Floquet engineering introduces temporal periodicity directly into the electronic Hamiltonian (Fig.~\ref{fig01}).
A periodic drive with frequency $\Omega$ creates nonequilibrium Floquet-Bloch states, characterized by quasienergy sidebands separated by integer multiples of $\hbar\Omega$~\cite{annurev-conmatphys-031218-013423,Phys.Rev.138.B979.1965,Phys.Rev.A.7.2203.1973}.
Although Floquet-Bloch states have been demonstrated in various materials using intense optical fields~\cite{Science.342.453.2013,Nat.Phys.12.306.2016,Nature.614.75.2023,Nat.Phys.16.38.2020,Nature.616.696.2023,Phys.Rev.B.111.L081106.2025,Nat.Phys.21.1093.2025,Nat.Phys.21.1100.2025}, in conventional pulsed implementations they are typically confined to the optical-pulse duration ($\sim 100$~fs).
Coherent phonons~\cite{ChemRev.94.157.2002,PhysRevB.45.768} offer an attractive alternative~\cite{Nano.Lett.18.1535.2018}, as they periodically modulate the electronic Hamiltonian and can retain phase coherence for tens of picoseconds.
Phonon-driven Floquet-Bloch states could therefore provide a substantially longer temporal window for probing and controlling nonequilibrium quantum phases, yet have not been experimentally observed.

The key challenge in identifying phonon-driven Floquet-Bloch states is to establish coherent quasienergy formation.
An externally applied laser field provides a phase-defined periodic drive.
In contrast, intrinsic collective excitations such as excitons and phonons may be populated with mutually uncorrelated phases and do not necessarily generate a coherent time-periodic field.
Thus, observing Floquet sideband-like spectral features alone does not entirely establish the presence of Floquet-Bloch states driven by intrinsic collective excitations.
A recent study has reported signatures consistent with exciton-driven Floquet physics~\cite{Nat.Phys.22.209.2026}, further highlighting the need for observables that directly access coherence.
Another challenge is that phonon-driven Floquet sidebands are separated only by the phonon energy $\hbar\Omega_{\mathrm{ph}}$, typically on the scale of a few to a few tens of meV, making them difficult to resolve spectroscopically.
Quantum-beat spectroscopy~\cite{LaserSpectroscopy1} provides a direct phase-sensitive probe: a coherent superposition of Floquet sidebands produces temporal oscillations whose frequency directly quantifies the quasienergy spacing, enabling the resolution of energy splittings below the nominal spectral resolution.

In this work, we demonstrate phonon-driven Floquet-Bloch states on graphene-covered Ir(111) using quantum-beat spectroscopy in time-resolved coherent photoemission~\cite{Science.277.1480.1997}.
We observe quantum beats arising from coherent interference between Floquet sidebands of image-potential states (IPSs).
The beat frequency matches that of an independently observed coherent phonon frequency, establishing quasienergy sidebands separated by $\hbar\Omega_{\mathrm{ph}}$ and in agreement with Floquet theory.
Moreover, the resulting states persist for one to two orders of magnitude longer than conventional light-driven Floquet-Bloch states.
These results provide the first experimental realization of phonon-driven Floquet-Bloch states and establish quantum-beat spectroscopy as a time-domain probe of Floquet matter.

\subsection*{Multiphoton photoemission of Graphene/Ir(111): Static characterization of image-potential states}

Graphene-covered Ir(111) is used to investigate phonon-driven Floquet effects.
The experimental setup, including sample preparation, is described in the Methods section.
Briefly, ultraviolet (UV; $\hbar\omega = 4.71$~eV) pump pulses are applied to excite coherent phonons and to populate electrons into the IPSs.
Infrared (IR; $\hbar\omega = 1.57$~eV) laser pulses are used to probe the electronic structure and dynamics of IPSs with a time delay $t$.
The normal-emission spectrum at $t = 0$ ps and the corresponding angle-resolved two-photon photoemission (2PPE) snapshot around the $\bar{\Gamma}$ point along the $\bar{\Gamma}$-$\bar{\mathrm{K}}$ direction are shown in Fig.~\ref{fig02}a.
In the normal-emission spectrum, three distinct peaks are observed; at the same energies, the angle-resolved 2PPE snapshot shows three parabolic bands around the $\bar{\Gamma}$ point.
These spectral features are assigned to the $n = 1 - 3$ IPSs~\cite{2012Phys.Rev.B.85.081402}.
Their peak energies, obtained from the Voigt-profile fits to the normal-emission spectrum, and their free-electron effective masses, obtained from the parabolic dispersion, are consistent with previous works~\cite{2012Phys.Rev.B.85.081402,Phys.Rev.B.112.L161408.2025}.
Furthermore, the high-energy side of the $n = 3$ peak exhibits an asymmetric shape, which we attribute to the emergence of the $n = 4$ state indicated by the red arrow in Fig.~\ref{fig02}a.
Since this weak spectral weight around the $n = 4$ state contains contributions from higher-order states whose energy positions are difficult to resolve individually, we denoted it as $n \geq 4$.
The background at a lower final-state energy originates from thermally excited electrons and reflects the Fermi-Dirac distribution tail, giving rise to an excess spectral weight beyond the $n = 1$ IPS line shape.

\subsection*{Quantum beat spectroscopy via phonon-driven Floquet sidebands}

\begin{figure}[!htbp]
	\centering
	\includegraphics[width=1\textwidth]{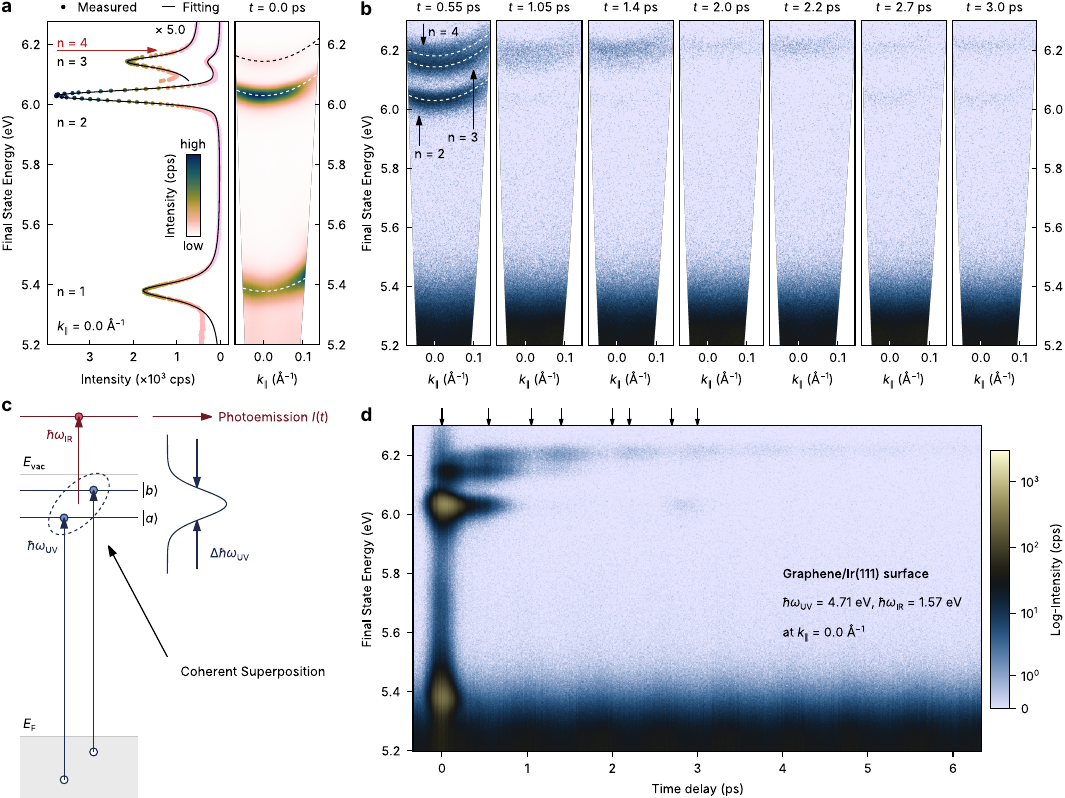}
    \begingroup
    \setstretch{1.2}
	\caption{\small\textbf{Phonon-driven Floquet-Bloch states on the IPSs of graphene-covered Ir(111) observed by quantum-beat spectroscopy.}
        \textbf{a,} Static 2PPE spectra of graphene-covered Ir(111) showing the IPSs measured with 1.57 and 4.71~eV photons.
        (left) The normal-emission spectrum.
        The high-energy asymmetry of the $n = 3$ peak indicates the emergence of the $n = 4$ state marked by the red arrow.
        At higher orders, adjacent IPSs become spectrally unresolved because their energy spacings are comparable to or smaller than the experimental resolution.
        (right) A linear-intensity map around the $\bar{\Gamma}$ point along the $\bar{\Gamma}$-$\bar{\mathrm{K}}$ direction.
        Dashed curves show parabolic fits to the $n = 1 - 3$ IPS dispersions.
        \textbf{b,} Log-intensity time- and angle-resolved 2PPE spectroscopy snapshots around the $\bar{\Gamma}$ point along the $\bar{\Gamma}$-$\bar{\mathrm{K}}$ direction.
        The selected delays follow the rapid decay of the resolved low-order IPSs and the oscillatory response in the unresolved region for $n \geq 4$.
        \textbf{c,} Energy schematic representing quantum beat spectroscopy in photoemission.
        The finite bandwidth $\Delta\hbar\omega_{\mathrm{UV}}$ of the pump pulse coherently excites neighboring states $| a \rangle$ and $| b \rangle$, and the IR pulse probes the temporal evolution of the superposition of these states.
        \textbf{d,} Delay-dependent photoemission intensity map at the $\bar{\Gamma}$ point, plotted as final-state energy versus time delay.
        Arrows mark the delays shown in panel (B), tracing the repeated enhancement and suppression of the unresolved $n \geq 4$ signal.
        A weaker oscillatory contrast is also visible around the $n = 3$ state.
    } \label{fig02}
    \endgroup
\end{figure}

Figure~\ref{fig02}b presents selected snapshots of the log-intensity time- and angle-resolved 2PPE spectra around the $\bar{\Gamma}$ point along the $\bar{\Gamma}$-$\bar{\mathrm{K}}$ direction, which provide an energy-momentum evolution of the unoccupied states after photoexcitation.
The delay sequence in Fig.~\ref{fig02}b shows that the resolved low-order IPSs ($n = 1$, $2$, $3$) decay rapidly, whereas the unresolved region for $n \geq 4$ remains visible for several picoseconds.
The noticeable population decay in these states is monotonic, with no significant dependence on the observed momentum.
Meanwhile, the $n \geq 4$ intensity exhibits clear oscillations as a function of delay time; it is stronger at $t = 0.55$, $1.4$, and $2.2$ ps and weaker at $t = 1.05$, $2.0$, and $2.7$ ps.
These observations indicate that the high-order IPS region is a clear spectral window in which an oscillatory 2PPE component is identifiable.

These oscillatory components can be described using the standard formulation of quantum-beat spectroscopy~\cite{Science.277.1480.1997}.
In the simplest case, where two excited states $| a \rangle$ and $| b \rangle$ are coherently accessible from a common initial state because the pump bandwidth exceeds their energy separation $E_b - E_a$ (Fig.~\ref{fig02}c), the pump prepares a coherent superposition,

\begin{equation}
    | \Psi (t) \rangle =
    c_a (t) e^{-i E_a t / \hbar} | a \rangle +
    c_b (t) e^{-i E_b t / \hbar} | b \rangle,
\end{equation}

\noindent\ where the coefficients $c_a (t)$ and $c_b (t)$ include the population decay and transition matrix elements.
The time evolution of the corresponding 2PPE intensity $I(t)$ is given by

\begin{equation}
    I(t) \propto % abs( Psi (t) )^(2) prop
    c_a^2 (t) + c_b^2 (t) +
    2 c_a (t) c_b (t) \cos \left[ ( E_b - E_a ) t / \hbar \right].
\end{equation}

\noindent\ This contains an interference term in addition to the decay terms.
Therefore, the beat frequency from the interference term provides the energy separation extracted from the delay-dependent spectra.

Figure~\ref{fig02}d shows a continuous energy-delay map at the $\bar{\Gamma}$ point to highlight the quantum beat dynamics, with arrows marking the delay times selected for the snapshots in Fig.~\ref{fig02}b.
Around the pump-probe overlap, the resolved low-order IPS features appear strongly, followed by order($n$)-dependent population dynamics.
The $n = 1$ signal decays rapidly with a smooth population envelope.
In parallel with these rapid decays, the quantum beat feature is clearly observed in the unresolved $n \geq 4$ region and persists for several picoseconds.
The arrow positions directly trace this oscillation: the $n \geq 4$ signal is enhanced near $t = 0.55$, $1.4$, and $2.2$ ps, corresponding to a frequency of 1.3 THz (5.4 meV).
The validity of this frequency assignment is confirmed by the Fourier analysis described later.
The observed beat energy of 5.4 meV agrees with the energy of a phonon mode previously identified in the same system by helium atom scattering (HAS)~\cite{Carbon.133.31.2018} or high-resolution electron energy-loss spectroscopy (HREELS)~\cite{andp.201400091} measurements, which has been assigned to the moir\'{e}-induced localized out-of-plane acoustic phonon of the graphene layer.
The correspondence between the 5.4 meV quantum-beat energy and the ladder-like Floquet quasienergy spacing provides strong evidence that the observed oscillatory quantum interference arises from phonon-driven Floquet sidebands.

Furthermore, a weaker oscillatory contrast is also observed around the $n = 3$ state.
At $t = 0.55$ ps, indicated by the second arrow, the $n = 3$ and $n \geq 4$ regions exhibit enhanced intensity, and the subsequent contrast reveals synchronized oscillation across the entire IPS linewidth.
In comparison, the corresponding feature for the $n = 2$ state appears only as a shoulder structure at $t = 0.55$ ps rather than as a clearly resolved oscillation pattern, due to the shorter lifetime of the $n = 2$ state.
The shared timing among these IPSs indicates that the same ladder-like quasienergy structure is formed coherently across all $n \geq 2$ IPSs, further reinforcing the phonon-driven Floquet picture as the origin of the quantum beat.
The multi-picosecond persistence of this phonon-energy quantum beat is consistent with the dressing of electron states by a coherent lattice vibration, providing time-domain evidence for phonon-driven Floquet-Bloch sidebands.

\subsection*{Coherent phonon signature in time-resolved multi-photon photoemission}

An independent, coherent lattice response is also observed in the same delay-energy map (Fig.~\ref{fig02}d).
Specifically, in a spectrally isolated final-state-energy window near the Fermi-Dirac distribution tail of the photoemission spectrum, a small but distinct intensity oscillation is observed.
This signature is consistent with the well-known interpretation of coherent phonons in time-resolved photoemission spectroscopy, where the coherent lattice motion modulates the electronic binding energy, inducing oscillations in the photoemission intensity~\cite{PhysRevLett.97.067402,RevModPhys.96.015003,ProgSurfSci.100.100795.2005}.
Regarding the IPSs, the intensity oscillations we have identified as the quantum beat do not support this interpretation, as no binding-energy modulation is observed.
In the current measurements, the oscillation persists beyond the maximum delay time of 6~ps, consistent with the long lifetime of the coherent phonon.

The probe-photon energy (1.57~eV) links the oscillations to a three-photon photoemission process.
As described later, the frequency of the intensity oscillation was determined to be $5.40 \pm 0.03$~meV.
This oscillation frequency matches the quantum-beat frequency observed in the $n = 3$ and $n \geq 4$ IPSs, while its spectral separation from the IPS windows renders it an independent reference for pump-driven surface dynamics.
We therefore assign this 5.40~meV oscillation to a coherent phonon and use it as an internal clock to test the phonon-driven Floquet interpretation described below.
The quantitative agreement between this internal lattice clock and the quantum-beat frequency provides direct evidence for the realization of phonon-driven Floquet-Bloch states, demonstrating that the periodic lattice distortion directly drives the sideband modulation.

\begin{figure}[!htbp]
	\centering
	\includegraphics[width=1\textwidth]{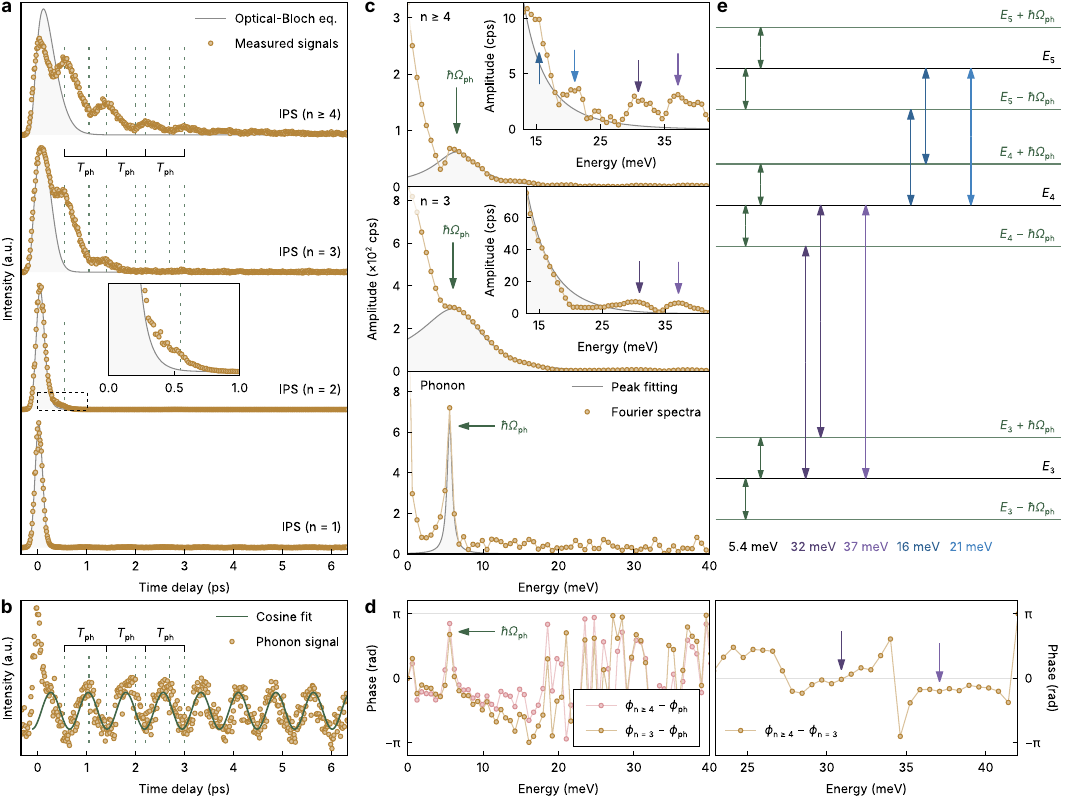}
    \begingroup
    \setstretch{1.2}
	\caption{\small\textbf{Time-domain and Fourier analysis of IPS quantum beats.}
        \textbf{a,} Integrated delay traces for the $n = 1$, $2$, $3$, and unresolved $n \geq 4$ IPS contributions.
        Solid curves show density-matrix fits representing the population envelopes.
        The $n = 1$ and $n = 2$ traces are dominated by population relaxation, while the $n = 3$ trace shows weak shoulder-like modulations, and the unresolved $n \geq 4$ transient exhibits a pronounced quantum beat.
        The inset highlights the shoulder structure in the $n = 2$ trace.
        \textbf{b,} Integrated delay trace of the Fermi-Dirac distribution tail intensity in the photoemission spectrum, used as an independent coherent-phonon channel.
        The fit after the prompt response gives a coherent oscillation at 1.30 THz, corresponding to an energy of 5.40~meV.
        \textbf{c,} Fourier amplitude spectra of the coherent-phonon, $n = 3$, and $n \geq 4$ traces.
        A common low-energy component near the phonon energy appears in all three spectra.
        Insets show higher-energy components near 32 and 37 meV, corresponding to one-phonon-shifted and parent inter-IPS coherence pathways, respectively.
        \textbf{d,} Fourier phase analysis using the coherent-phonon component as the phase reference.
        The $n = 3$ and $n \geq 4$ IPS beats show the same phase relation to the coherent phonon, indicating phase-locked quantum beats driven by the same coherent lattice motion.
        \textbf{e,} Energy diagram of the $n=3$, $n=4$ and $n=5$ IPSs and their phonon-driven Floquet sidebands.
        Coherent superposition between the parent IPSs ($E_4 - E_3$ and $E_5 - E_4$) and the corresponding sideband states gives rise to the observed quantum beat components, in addition to the coherent superposition between adjacent sidebands (5.4 meV).
    } \label{fig03}
    \endgroup
\end{figure}

\subsection*{Fourier analysis of quantum beat: Amplitude and phase}

To quantify the time-domain oscillations identified in Fig.~\ref{fig02}, we integrate the photoemission intensity within defined energy windows in the energy-delay map, yielding the time-domain traces shown in Figs.~\ref{fig03}a and \ref{fig03}b.
For the IPSs, energy windows comparable to the probe bandwidth ($\sim 20$~meV) centered at their respective peak positions, and for the coherent-phonon channel, a distinct final-state-energy interval ($5.2 - 5.24$ eV) is selected.
Figure~\ref{fig03}a shows the IPS transients together with density-matrix fits based on the optical Bloch equations~\cite{PhysRevLett.76.535,UEBA2007193}.
As described above, the $n = 1$ traces are dominated by population relaxation, whereas the $n = 2$ trace shows weak shoulder-like deviations, the $n = 3$ trace shows a weak oscillation, and the unresolved $n \geq 4$ region exhibits a pronounced oscillatory modulation.
The weak features observed in the $n = 2$ and $n = 3$ transients align with the maxima of the $n \geq 4$ oscillation, consistent with the in-phase contrast observed in the energy-delay map (Fig.~\ref{fig02}d).
No additional peak or shoulder structure is observed in the $n = 1$ transient, presumably attributed to the significantly shorter lifetime of the $n = 1$ IPS (see Fig.~\ref{figS1}).

Figure~\ref{fig03}b shows the spectrally distinct coherent-phonon channel, where the integrated intensity exhibits a prompt response followed by a single dominant oscillation.
A Fourier transform of the data shown in Fig.~\ref{fig03}b for $t \geq 0.3$ ps reveals a coherent phonon energy of $5.40 \pm 0.03$~meV (Fig.~\ref{fig03}c).
We then perform the same Fourier analysis on these time-domain traces to compare the frequency components of the coherent phonon, the $n = 3$, and the unresolved $n \geq 4$ IPSs.
The Fourier amplitude spectra in Fig.~\ref{fig03}c show a common low-energy component in all three traces.
Peak fitting gives $5.46 \pm 0.3$ meV for the $n = 3$ transient and $5.30 \pm 0.3$ meV for the $n \geq 4$ transient.
The shared Fourier component near 5.40 meV therefore establishes that the IPS quantum beats and the independently measured coherent phonons carry the same energy quantum.

The Fourier phase analysis further examines whether these oscillations are locked to a common coherent motion.
Figure~\ref{fig03}d compares the phases of the $n = 3$ and $n \geq 4$ IPS quantum beats with that of the coherent-phonon signal extracted from the Fermi-Dirac-tail intensity oscillation.
Both quantum beats are essentially shifted by $\pi$ relative to the coherent-phonon signal, indicating that they share the same phase relationship with respect to the coherent phonon.
This common phase relation also explains the in-phase alignment between the weak shoulders in the $n = 3$ transient and the pronounced maxima in the $n \geq 4$ transient.
The phase analysis, therefore, converts the energy matching in the amplitude spectra into temporal evidence that both IPS quantum beats are locked to the same coherent lattice oscillation.
The finite phase difference between the coherent phonon and the IPS response is consistent with the Floquet picture, in which periodic modulation of the IPS Hamiltonian generates phase-locked sideband components~\cite{Nano.Lett.18.1535.2018,Nature.616.696.2023}.

Additionally, the higher-energy Fourier peaks reveal coherence between neighboring IPS-derived components and their phonon-shifted Floquet sidebands.
Specifically, the $n = 3$ and $n \geq 4$ spectra share two peaks in the higher-energy region (Fig.~\ref{fig03}c, insets): one at around 32 meV and the other at around 37 meV.
We find that the lower-energy peak at 32 meV is consistent with the value of $E_4 - E_3 - \hbar\Omega_{\mathrm{ph}}$, indicating quantum interference between the Floquet sideband of the $n = 3$ IPS and the static $n = 4$ IPS, or vice versa, as illustrated in Fig.~\ref{fig03}e.
Meanwhile, the higher-energy value is likely associated with the static energy difference between the $n = 3$ and $n = 4$ IPS levels ($E_4 - E_3$), suggesting coherent interference between neighboring IPS components.
Notably, the phases of these higher-energy components in Fig.~\ref{fig03}d remain in phase across the $n = 3$ and $n \geq 4$ spectra, indicating that these cross-IPS coherences arise from the same pump-driven temporal event.
Together, the lower- and higher-energy Fourier peaks reveal that a coherent phonon induces intra-IPS Floquet sideband quantum beats at $\hbar\Omega_{\mathrm{ph}}$ and induces quantum coherence between neighboring IPS Floquet sidebands.
Likewise, as shown in the inset of the top panel of Fig.~\ref{fig03}c, a characteristic pair of features, consisting of a shoulder structure at around 16 meV and a small peak at around 21 meV, appears in the spectrum for $n \geq 4$, but is absent in the $n = 3$ spectrum.
These energies are consistent with $E_5 - E_4 - \hbar\Omega_{\mathrm{ph}}$ and $E_5 - E_4$.
With the pump bandwidth of approximately 34 meV, the absence of higher-energy components is consistent with the experimental sensitivity.

Finally, we address photoemission-related replica mechanisms that can mimic Floquet sidebands.
In light-driven Floquet engineering, Volkov states that the emitted photoelectrons are dressed by an optical field, producing final-state replicas that must be distinguished from the Floquet-Bloch sidebands.
Such Volkov states are ruled out as the origin of the observed 5.40 meV component because final-state optical dressing produces photon-energy sidebands that are tied to the temporal overlap of the optical pulse.
A more relevant possibility is conventional electron-phonon interaction, which can produce an inelastic-interaction satellite in photoemission spectra~\cite{Nature515.7526}.
Since these features are spaced by the phonon energy, energy-domain analysis alone cannot uniquely identify phonon-driven Floquet dressings.
However, these replicas arise from inelastic scattering during photoemission~\cite{PhysRevLett.120.237001,PhysRevLett.95.207601,PhysRevB.88.224301} and do not involve a phase-locked coherent superposition of IPS-derived components.
Such mechanisms, therefore, cannot account for the observed quantum beats in the time domain, their several-picosecond persistence, and their fixed phase relation to the coherent-phonon channel.

\subsection*{Conclusion and outlook}

While the present experiment establishes phonon-driven Floquet dressing through time-domain quantum beats, several expected signatures remain beyond the scope of the present observables.
The emergence of an energy gap and its dependence on pump polarization are well-established experimental signatures of light-driven Floquet-Bloch states.
In contrast to conventional Floquet studies based on gap formation, the present study identifies Floquet dressing through the coherence of IPS-derived wave functions, manifested as phonon-energy quantum beats in the time domain.
In the present IPS system, the relevant dispersions are nearly parallel in momentum space due to their similar effective masses; thus, the avoided-crossing feature associated with Floquet hybridization is not clearly observed.
Systems with more pronounced dispersion differences would allow quantum beat spectroscopy to probe Floquet gap formation through momentum-dependent modification or suppression of the beating structure at specific momenta.
Polarization control offers a crucial avenue in light-driven Floquet dressing studies.
Specifically, polarization dependence serves as a key criterion for distinguishing intrinsic Floquet-Bloch signatures from final-state Volkov replicas~\cite{Nat.Phys.12.306.2016}.
Furthermore, the use of circularly polarized light can break the time-reversal symmetry and induce additional gap features~\cite{Science.342.453.2013}.
By analogy, phonon-driven Floquet engineering calls for control over phonon polarization, including linearly polarized lattice motion and chiral phonons~\cite{PhysRevLett.115.11550}.
Finally, the present work focuses on a single-phonon-mode-driven Floquet scenario, whereas a multimode-phonon process can lead to complex Floquet quasienergy structures.
Such multimode Floquet engineering may therefore be particularly advantageous in phonon-driven systems, where the relatively long coherence time of lattice motion can sustain a more stable periodic drive beyond the excitation laser pulse duration.

We have demonstrated experimental evidence for phonon-driven Floquet sidebands on graphene/Ir(111) using quantum-beat spectroscopy.
The $n = 3$ and unresolved $n \geq 4$ IPSs exhibit quantum beats at 5.40 meV, consistent with an independently observed coherent-phonon signal.
Fourier amplitude and phase analyses further show that these beats share the phonon energy and remain phase-locked to the coherent lattice motion, establishing their origin as phonon-driven Floquet dressing.
These results provide direct evidence that coherent phonons can generate Floquet sidebands and offer a viable route toward Floquet engineering using material-driven phonon fields rather than optical fields, with substantially longer lifetimes that enable experimentally accessible regimes beyond conventional light-driven platforms.
As temporal coherence is an indispensable property of Floquet-Bloch states regardless of the driving source, quantum beat spectroscopy provides a powerful and broadly applicable probe of Floquet physics.
More broadly, this work demonstrates that time-resolved photoemission, combined with phase-resolved Fourier analysis across multiple electronic states, can resolve temporal relationships among the lattice drive, Floquet dressing, and appearance of dressed-state features, even when energy-domain resolution is insufficient.
These capabilities establish a framework for exploring and engineering material-driven Floquet states in quantum materials beyond the limits of conventional light-driven platforms.

%%=============================================%%
%% For presentation purpose, we have included  %%
%% \bigskip command. Please ignore this.       %%
%%=============================================%%

\backmatter

\bmhead{Acknowledgements}

R.A. would like to thank Dr. J. Inoue and Dr. T. Uchihashi of NIMS for fruitful discussions.
\paragraph*{Funding:}
This work was financially supported by JPSJ KAKENHI, Japan (grant No. 26K01393, 26K21854).
It was also partially supported by the National Science and Technology Council of Taiwan (grant No. 113-2112-M-A49-020-MY3, 113-2628-M-A49-006-MY3, 114-2124-M-A49-003, 114-2923-M-A49-001-MY2) and
the NSTC T-Star Center Project: Future Semiconductor Technology Research Center (grant No. 115-2634-F-A49-009).
\paragraph*{Author contributions:}
Y.-C.T. performed the photoemission experiments including sample preparation, analyzed the data, and wrote the manuscript with input from all authors.
C.-W.L. advised on the design of the optical setup.
N.T., H.I., and R.A. performed theoretical (DFT) calculations.
N.T. and C.-L.L. performed STM measurements.
R.A. supervised the project.
\paragraph*{Competing interests:}
The authors declare that they have no competing interests.
\paragraph*{Data, code, and materials availability:}
All data are available in the main text or the supplementary materials.
The material used in this study was prepared as described in the supplementary materials.


\begin{appendices}

% Figures, tables, equations and pages in the supplement are numbered S1, S2 etc.
\renewcommand{\figurename}{Extended Data Fig.}
\renewcommand{\thefigure}{\arabic{figure}}
\renewcommand{\theHfigure}{ext.data.\arabic{figure}}
\setcounter{figure}{0}

\clearpage{}
\subsection*{Methods}
\subsubsection*{Sample preparation}

\begin{figure}[!htbp]
	\centering
	\includegraphics[width=0.67\textwidth]{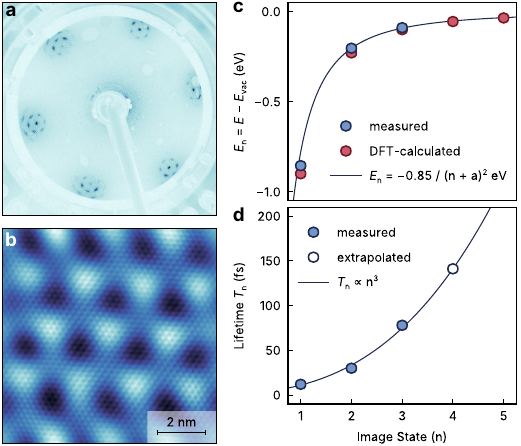}
    \begingroup
    \setstretch{1.2}
	\caption{\small
        \textbf{a,} Low-energy electron diffraction (LEED) pattern of the graphene-covered Ir(111) surface.
        Sharp diffraction spots from Ir, C, and the surrounding moir\'{e} satellites are observed, indicating high structural quality and homogeneity of the sample.
        \textbf{b,} Scanning tunneling microscope (STM) image showing the formation of a long-period moir\'{e} superlattice due to the lattice mismatch between graphene and Ir(111) surface, in addition to the carbon atoms of graphene.
        \textbf{c,} Energy positions of the image-potential states (IPSs) at the $\bar{\Gamma}$ point.
        The filled blue circles represent the data shown in Fig.~\ref{fig02}a.
        The red circles show the DFT-calculated values.
        The curve represents a fit to a Rydberg-like series, $E_n = -0.85 / (n + a)^2$ eV with $a = 0.04$.
        \textbf{d,} Energy relaxation time evaluated from the density-matrix fit based on the optical Bloch equation.
        The dashed curve represents the $T \propto n^3$ fit.
    } \label{figS1}
    \endgroup
\end{figure}

Graphene was grown on Ir(111) by chemical vapor deposition, using the Ir substrate's catalytic activity to decompose ethene \cite{New.J.Phys.11.023006.2009}.
The Ir(111) surface was cleaned by repeated Ar-ion sputtering (1.5 keV, 30 min), followed by annealing at 1370 K for 30 min.
Graphene was synthesized via a well-established two-step method involving temperature-programmed growth and chemical vapor deposition.
In the first step, ethene (C\textsubscript{2}H\textsubscript{4}) was adsorbed onto the Ir(111) surface at room temperature, and then the sample was flashed to 1370~K for 60~s to form graphene islands.
Subsequently, the sample was exposed to ethene gas ($2 \times 10^{-8}$ mbar) at 1100~K for 300~s.
Surface cleanliness was confirmed by sharp diffraction spots in low-energy electron diffraction (LEED) patterns (Extended Data Fig.~\ref{figS1}a).
The clear and sharp satellite spots originating from the moir\'{e} pattern observed in LEED indicate uniform graphene coverage without rotational domains.
The surface morphology was further characterized by scanning tunneling microscopy (STM), which revealed a well-ordered moir\'{e} superlattice with a periodicity of approximately 2.5~nm, consistent with the lattice mismatch between graphene and Ir(111) (Extended Data Fig.~\ref{figS1}b).

\subsubsection*{Experimental details}

Our apparatus for time- and angle-resolved multiphoton photoemission spectroscopy measurements consists of a Ti:sapphire laser system and a hemispherical electron energy analyzer, both placed in an ultrahigh-vacuum chamber with a base pressure of $2 \times 10^{-10}$ mbar.
The incident laser pulses are generated by an 80 MHz Ti:sapphire oscillator, which produces 790 nm (1.57 eV) fundamental pulses with a duration of approximately 120 fs.
Frequency-tripled ultraviolet (UV) pulses at 4.71 eV from the oscillator output serve as the pump to excite the sample, while the fundamental infrared (IR) pulses serve as the probe to measure electrons from the transiently populated states.
The p-polarized pump and probe pulses are incident at 45$^{\circ}$ from the sample surface and are focused onto the sample with a spot size of approximately 100~$\mathrm{\mu m}$ in diameter.
The emitted electrons are measured at room temperature using a hemispherical analyzer equipped with a two-dimensional detector.
The total energy resolution is maintained at approximately 20 meV, determined by the probe pulse bandwidth and the analyzer resolution.

\subsubsection*{Calculation procedure}

To evaluate the energy levels of the IPSs on graphene/Ir(111), we employed the following two-step first-principles calculation procedure.
In the first step, the surface geometry was optimized; in the second, the electronic structure was calculated using the optimized geometry.
The structural optimization was performed using the VASP code~\cite{VASP1,VASP2} with the projector-augmented-wave (PAW) method~\cite{PhysRevB.50.17953}.
The local density approximation (LDA) was employed as the exchange-correlation functional.
The surface structure was modeled as a slab consisting of a ($10 \times 10$) graphene layer on four Ir(111) layers, each a ($9 \times 9$) supercell, as determined from LEED and STM observations.
All atoms except for those in the bottom Ir layer were relaxed until the residual forces were less than 0.01~eV/\AA.

To calculate the energy levels of the IPSs, we then use another independent computational code that combines the embedded Green's function method with the full-potential linearized augmented plane-wave methods~\cite{2001Phys.Rev.B.63.165409,1981J.Phys.C:Solid.State.Phys.14.3795,2006SinghPlanewavesPseudopotentialsLAPW,PhysRevB.102.195425}.
The density functional theory (DFT) potential $\bar{V}_{\mathrm{eff}}$ was smoothly matched to a classical image potential~\cite{1995Prog.Surf.Sci.50.149} to describe the IPS.
The calculated work function was 4.846 eV, which is higher than the experimentally determined value of 4.66 eV.
The IPS binding energies were defined as the difference between the work function and the IPS energies measured from the Fermi level.

\subsubsection*{Basic Characterization of Image Potential States on Graphene/Ir(111)}

From the low-energy cutoff of the one-photon photoemission spectrum, we determine a work function of 4.66 eV for graphene/Ir(111).
This calibration is essential because image-potential-state binding energies are referenced to the vacuum level, and the graphene overlayer modifies that alignment relative to bare Ir(111).
The reduced work function highlights graphene's role in reshaping the near-surface potential and enabling efficient access to higher-order image-potential states.

Peak positions extracted at the $\bar{\Gamma}$ point are well described by a Rydberg-like series, $E_n = -0.85 / (n + a)^2$ eV, with a small quantum defect ($a = 0.04$).
Extended Data Figure~\ref{figS1}c also shows the corresponding values evaluated from the DFT calculations.
These results establish the intrinsic IPS energy-level structure that serves as the reference for the time-domain analysis presented in the main text.

As shown in Fig.~\ref{fig03}A, a relaxation-only density-matrix population model provides a baseline for interpreting the delay-dependent IPS dynamics.
This model captures the dominant prompt population and decay in the resolved low-order states, yielding an energy relaxation time of $T_1 = 12.0 \pm 2.5$ fs for the $n = 1$ state.
For the $n = 2$ ($T_2 = 30.0 \pm 4.0$) and $n = 3$ ($T_3 = 78.0 \pm 9.0$ fs) states, however, additional late-time structures emerge, including a distinct increase in intensity near 0.55 ps that is not reproduced by the relaxation-only description.
The unresolved higher-order $(n \geq 4)$ trace shows the strongest deviation: while its overall decay timescale ($T_4 \approx 141$ fs) is reasonably estimated, the pronounced oscillatory modulation remains unexplained by the population model.
These deviations indicate the presence of additional coherent dynamics beyond a simple relaxation-only description.




\end{appendices}


\bibliography{sn-bibliography}% common bib file
%% \input{sn-article.bbl}

\end{document}